\documentclass{aastex}
\usepackage{emulateapj5}

\def\lesssim{\mathrel{\hbox{\rlap{\hbox{\lower4pt\hbox{$\sim$}}}\hbox{$<$}}}}
\def\gtrsim{\mathrel{\hbox{\rlap{\hbox{\lower4pt\hbox{$\sim$}}}\hbox{$>$}}}}

\shorttitle{Bar Formation in Magnetic Clouds}
\shortauthors{Nakamura \& Li}

\begin{document}

\title{Nonaxisymmetric Evolution of Magnetically Subcritical
Clouds: Bar Growth, Core Elongation, and Binary Formation}

\author{Fumitaka Nakamura}
\affil{Faculty of Education and Human Sciences, Niigata University,
8050 Ikarashi-2, Niigata 950-2181, Japan, and Astronomy Department,
University of California, Berkeley, Berkeley, CA 94720} 
\author{Zhi-Yun Li}
\affil{Astronomy Department, University of Virginia, Charlottesville,
VA 22903}

\begin{abstract}
We have begun a systematic numerical study of the nonlinear growth
of nonaxisymmetric perturbations during the ambipolar
diffusion-driven evolution of initially magnetically subcritical
molecular clouds, with an eye on the formation of binaries,
multiple stellar systems and small clusters. In this initial
study, we focus on the $m=2$ (or bar) mode, which is shown to be
unstable during the dynamic collapse phase of cloud evolution after
the central region has become magnetically supercritical. We
find that, despite the presence of a strong magnetic field, the
bar can grow fast enough that for a modest initial perturbation
(at $5\%$ level) a large aspect ratio is obtained during the
isothermal phase of cloud collapse. The highly elongated bar
is expected to fragment into small pieces during the subsequent
adiabatic phase. Our calculations suggest that the strong magnetic
fields observed in some star-forming clouds and envisioned in
the standard picture of single star formation do not necessarily
suppress bar growth and fragmentation; on the contrary, they may
actually promote these processes, by allowing the clouds to
have more than one (thermal) Jeans mass to begin with without
collapsing promptly. Nonlinear growth of the bar mode in a direction
perpendicular to the magnetic field, coupled with flattening
along field lines, leads to the formation of supercritical cores
that are triaxial in general. It removes a longstanding objection
to the standard scenario of isolated star formation involving
subcritical magnetic field and ambipolar diffusion based on the
likely prolate shape inferred for dense cores. Continuted growth
of the bar mode in already elongated starless cores, such as
L1544, may lead to future binary and multiple star formation.

\end{abstract}


\section{Introduction}
\label{sec:introduction}

A basic framework has been developed for the formation of low-mass
stars in relative isolation \citep{FShu87}. It involves
gradual condensation of dense cores from strongly magnetized,
background molecular clouds through ambipolar diffusion, followed
by dynamic core collapse to form stars. Quantitative studies based
on this by now ``standard'' scenario have been carried out by many
authors, with increasingly sophisticated input physics
\citep{TMouschovias99}. However, with few exceptions
\citep{RInde00, ABoss00}, axisymmetry has been adopted. The adopted symmetry
precludes a detailed investigation of cloud fragmentation, generally
thought to be a necessary step in the formation of binary and
multiple stellar systems; it is in such systems that most stars are
found. We seek to improve this situation by removing the restriction
of axisymmetry and systematically investigating the effects of
the strong magnetic fields envisioned in the standard picture
of single star formation on fragmentation and their implications
on binary and multiple star formation.

Fragmentation of nonmagnetic clouds has been studied extensively
over the years, mostly through numerical simulations \citep[and
references therein]{PBodenheimer00}.
In the canonical case of a
spherical cloud, the fragmentation is mainly controlled by the ratios
of the cloud thermal and rotational energy to the gravitational
energy, $\alpha$ and $\beta$, although the distributions of density
and angular momentum also play a role. Molecular line observations
\citep{AGoodman93} suggest that star-forming cores of molecular
clouds are far from being rotationally supported, with a typical
(low) value of $\beta \sim 0.02$. For such slowly rotating cores,
the criterion for fragmentation is roughly $\alpha \lesssim 0.5$,
if the cores are idealized as spheres of uniform density and rigid
rotation \citep{TTsuribe99}. More realistic, centrally peaked density
distributions would lower the critical $\alpha$ for fragmentation. 
Such distributions are commonly inferred in the well-observed starless
cores such as L1544 \citep{PAndre00, NEvans01}. The central
concentration, if taken as an initial condition, tends to make
fragmentation more difficult \citep{ABoss93}.

Fragmentation of strongly magnetized clouds with mass-to-flux ratios
below the critical value $(2\pi)^{-1}G^{-1/2}$ (i.e., subcritical)
is expected to be quite different \citep{DGalli01}. 
Through magnetic braking, the
field controls the angular momentum evolution of the cloud, until
a supercritical dense core forms \citep{SBasu94}; it
thus determines the value of $\beta$ of the core. More importantly,
the field can provide most of the cloud support against self-gravity
and thus allow a cloud to have an arbitrarily small thermal energy
compared to the gravitational energy and still be in a mechanical
equilibrium initially. The low value of $\alpha$ is a key ingredient
of fragmentation. On the other hand, the presence of a strong magnetic
field tends to stifle cloud fragmentation, if the field and matter
are well coupled \citep{GPhillips86}. Indeed, for a frozen-in
subcritical field, fragmentation is suppressed altogether \citep{TNakano88}.
Fragmentation can in principle resume in the part of cloud
that has become magnetically supercritical \citep{ABoss00}, through
ambipolar diffusion. It is however difficult to predict {\it a priori}
how the fragmentation should proceed, if at all, in a cloud that
is only {\it partially} coupled to a strong magnetic field.

\citet{WLanger78} showed through linear analysis that, in a lightly
ionized medium such as a molecular cloud, the Jeans instability
is not completely inhibited no matter how strong the magnetic field
is; rather, it grows on an ambipolar diffusion, rather than dynamic, 
timescale. 
\citet{ZLi01} followed the nonlinear evolution of a set of magnetically
subcritical, super-Jeans mass clouds assuming
axisymmetry and found that either a dense, supercritical core or
ring forms as a result of ambipolar diffusion. The supercritical
ring is expected to fragment readily into a number of dense
cores, producing perhaps a small group of stars. The ring breakup
will be studied elsewhere. In this Letter, we shall concentrate on
the more subtle problem of the fragmentation of magnetized single
cores, with an eye on forming binaries and multiple stellar systems.
As is common with numerical studies of cloud fragmentation and
binary formation, we focus on the growth of $m=2$ (or bar) mode
\citep[e.g.,][]{FNakamura97, PBodenheimer00}.
In \S~\ref{sec:models} we describe our formulation of the problem
of nonaxisymmetric cloud evolution. Numerical
results on bar formation are presented in \S~\ref{sec:results},
and their implications on binary and multiple star formation
as well as the observed elongation of dense cores are discussed
in \S~\ref{sec:conclusions}.

\section{Problem Formulation}
\label{sec:models}

An ordered, subcritical magnetic field allows cloud material to settle
along field lines into a disk-like geometry. We assume that the disk
is in hydrostatic equilibrium in the vertical direction,
and adopt the thin-disk approximation \citep{ZLi01}. 
The cloud evolution is followed in the $x$-$y$ plane 
of a Cartesian coordinate system $(x,y,z)$. Outside
the disk, the magnetic field is assumed to be current free and
uniform far from the cloud. An isothermal equation of state is adopted
(unless noted otherwise), so that the pressure and surface density are
related through $P = c_s^2 \Sigma $, where $c_s$ is the (effective)
isothermal sound speed. 
The governing equations
are the nonaxisymmetric versions of equations (1) through (7)
of \citet{ZLi01}.

The initial conditions for star formation are not well determined
either observationally or theoretically. Following 
\citet{SBasu94} we prescribe an axisymmetric reference state
for our model clouds with the distributions of mass and magnetic field
given by
\begin{equation}
\Sigma _{\rm ref} (x,y) = \frac{\Sigma_{\rm 0,ref}}
{\left[1+r^2/r_0^2\right]^{3/2}}\ \ \ {\rm and} \ \ \
B_{z, \rm ref} (x,y) = B_{z, \infty},
\end{equation}
where $r= (x^2+y^2)^{1/2}$. 
The background field strength $B_{z, \infty}$
is characterized by the dimensionless
flux-to-mass ratio $\Gamma_0=B_{z, \infty}/[2\pi G^{1/2} \Sigma_{0, \rm
ref}]$. The reference clouds are not in an equilibrium state and are
allowed to evolve into one with magnetic field frozen-in, before
ambipolar diffusion is turned on at time $t=0$. At $t=0$, we impose
on top of the equilibrium surface density $\Sigma_0(x,y)$ an
$m=2$ perturbation of relative amplitude $A$, 
$\Sigma (x,y)= \Sigma_0(x,y)\ [1 + A\; \cos(2 \phi)]$,
where $\phi$ is the azimuthal angle measured from the $x$-axis. A slow
rotation is also added to the cloud at this time, according to the
prescription
\begin{equation}
V_\phi (x,y)={ 4 \; \omega \; r \over r_0
	+\sqrt{r_0^2+r^2} } c_{ms}  ,
\end{equation}
where $c_{ms}=c_s (1+\Gamma_0^2)^{1/2}$ is essentially the magnetosonic 
speed, and $\omega$ measures the rotation rate.
The subsequent, ambipolar diffusion-driven evolution of the perturbed
cloud is followed numerically, subject to the condition of fixed
$\rho$ and $B_z$ at the cloud outer radius, taken to be twice the
characteristic radius $r_0$. Other choices of the reference state will 
be explored elsewhere.

Our calculations are carried out using an MHD code based on that of
\citet{FNakamura97}. The hydrodynamic part is solved using Roe's
 method. 
The equation for magnetic
field evolution has a form identical to that for mass conservation,
and is solved in the same manner. Both the magnetic and gravitational
potentials satisfy the Poisson equation, 
and 
are solved using a convolution method 
 \citep{RHockney81}.  We refine the computational grids whenever
the so-called Jeans condition is about to be violated
\citep{KTruelove97}, taking into account the magnetic effect on
Jeans condition. 
The computations are carried out using nondimensional
quantities. The units we adopted are $c_s$ for speed,
$\Sigma _{0,\rm ref}$ for surface density, $2\pi G\Sigma_{0,\rm ref}$
for gravitational
acceleration, and $B_{z,\infty}$ for magnetic field strength.
The units for length and time are, respectively, $L_0\equiv
c_s^2/(2\pi G\Sigma_{0,\rm ref})$ and $t_0\equiv c_s/(2\pi
G\Sigma_{0,\rm ref})$.

\section{Numerical Results}
\label{sec:results}

The results of our calculations are shown in two figures. In Figure 
\ref{fig:1},
we show the evolution of a ``benchmark'' cloud with the dimensionless
radius $r_0=7.5\pi$, initial flux-to-mass ratio 
$\Gamma_0=1.5$ and rotation rate $\omega=0.1$. 
The $m=2$ perturbation added has a
relative amplitude of ``merely'' $A=0.05$. During the initial,
quasi-static stage of evolution, a central core
condenses gradually out of the magnetically subcritical cloud,
with no apparent tendency for the mode to grow. Rather,
the iso-density contours appear to oscillate, changing the
direction of (slight) elongation along the $x$-axis in the disk 
plane to $y$-axis, as evident from panels (a) and (b).
(We confirmed that the disk continues to oscillate around the
equilibrium state in the absence of ambipolar diffusion.)
The period of the oscillation is more or less independent of $\omega$
as long as $\omega$ is small. 
After a supercritical core develops,
the contraction becomes dynamic, as seen from the velocity field
in panel (c). By this time, the $m=2$ mode has grown significantly,
resulting in a bar-like core at the center with an aspect ratio
of roughly 2. As the collapse continues, the bar growth remains
relatively modest [panel (d)] until the very end of the starless
collapse, when the growth rate increases dramatically [panel (e)].
Beyond a critical dimensionless surface density of $1.9\times 10^4$,
corresponding to a volume density of $n_H=10^{12}$~cm$^{-3}$, we
change the equation of state from isothermal to adiabatic, with
an index of $5/3$, to mimic the transition to optically thick
regime of cloud evolution. The density distribution and velocity
field around the time when an accretion shock begins to develop
are shown in panel (f). The bar bound by the shock is analogous
to the ``first'' core of spherical calculations. It has an aspect
ratio of about 13. This highly elongated ``first'' bar is expected
to break up into two or more pieces, as discussed further in
\S~\ref{sec:conclusions}.

\begin{figure}
\epsscale{0.6}
\plotone{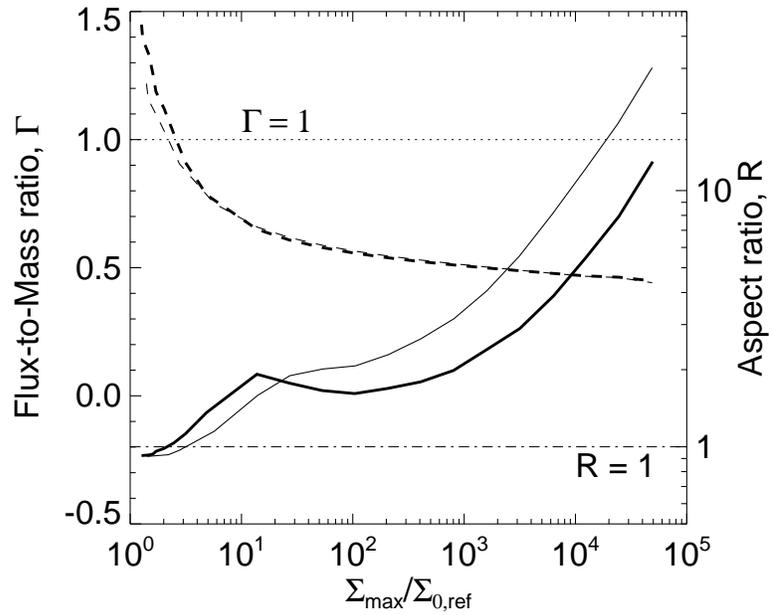}
\caption{
Evolution of the aspect ratio R of the bar ({\it solid lines})
and the flux-to-mass ratio $\Gamma$ (in units of the critical value)
at the surface density maximum
 ({\it dashed lines}).
Thick and thin lines are for models with
$(\Gamma_0, r_0, \omega)= (1.50, \, 7.5\pi, \, 0.1)$ and
$(\Gamma_0, r_0, \omega)= (1.25, \, 7.5\pi, \, 0.1)$, respectively.
For reference, lines of $\Gamma = 1$ and $R=1$ are also plotted.
\label{fig:2}}
\end{figure}

To discuss the bar growth more quantitatively, we plot 
in Figure \ref{fig:2}
the aspect ratio $R$ of the darkest region in Figure \ref{fig:1}
(with $\Sigma
\geq 10^{-1/2}\Sigma_{\rm max}$) as a function of the central
surface density $\Sigma_{\rm max}$, together with the flux-to-mass
ratio $\Gamma$ at the cloud center. It is clear that the $m=2$
mode grows little when the entire cloud remains subcritical (i.e.,
$\Gamma > 1$), as one might expect intuitively. As the central region
becomes more and more supercritical, the mode grows significantly,
producing a bar of aspect ratio $\sim2$. This ratio stays more or
less ``frozen''
during the subsequent evolution, presumably due to the onset of
rapid collapse, which leaves little time for the bar to grow. 
The flux-to-mass ratio does not change much either during this
period, also as a result of the rapid collapse which prevents 
the magnetic flux from leaking out. The fairly strong trapped 
field (with $\Gamma\gtrsim 0.5$) further hinders the bar growth, 
which picks up speed only towards the end of collapse before 
forming a singularity, as a result of the \citet{CLin65} instability.

We have studied other models with different initial parameters in
the range of $1.25\le \Gamma\le 2$, $5\pi\le r_0\le 10\pi$, 
$0\le \omega\le 0.1$, and $0.05 \le A \le 0.1$. 
We find that the bar grows more rapidly for 
a smaller $\Gamma_0$, larger $r_0$, larger $\omega$, or larger $A$. 
For example, if we choose $\Gamma_0 = 1.25$, $r_0 = 7.5 \pi$ 
(adopted by \citet{GCiolek00} for their axisymmetric 
model of the well-observed starless core L1544), 
$\omega=0.1$, and $A=0.05$, 
then the aspect ratio reaches 33 when the ``first'' bar forms, as 
shown in Figure \ref{fig:2}. Interestingly, a plateau region is also 
evident, where the aspect ratio is temporarily ``frozen'' at $R \sim 2$, 
as in the benchmark model. We find that the frozen value of the aspect
ratio depends weakly on the initial amplitude of perturbation,
which may have observable consequences as we discuss below.

\section{Discussion and Conclusion}
\label{sec:conclusions}

We have followed numerically the growth of $m=2$ mode during the
ambipolar-diffusion driven evolution of magnetically subcritical
clouds. Our main conclusion is that, despite the presence of the
strong magnetic field, a perturbation of modest amplitude can
grow nonlinearly into a highly elongated bar, which is expected
to fragment into small pieces gravitationally. Before discussing
fragmentation, we comment on the effects of bar growth on the
observed shapes of molecular cloud cores.

Dense cores of molecular clouds are intimately associated with star
formation, with roughly half of them already harboring infrared
sources \citep{CBeichman86}. The other half are thought to be
well on their way to star formation, with the ``starless'' phase
lasting for only a few dynamic times. The cores are observed
to have significant elongation, with a typical aspect ratio of 2
\citep{PMyers91}. Statistics of core elongation have been
interpreted as indicating most cores have prolate 3-dimensional
shape \citep{BRyden96, CCurry01}, although oblate cores formed as a result
of settling along magnetic field lines can have a projected
aspect ratio similar to that observed \citep{ZLi96}. Nonlinear
growth of the bar mode in a direction perpendicular to the field
lines modifies the core shape drastically, making it triaxial
in general. Some evidence for the triaxial nature of cores
has been marshalled by \citet{SBasu00}.
We propose that it is due to the significant bar growth during
the transition period when $\Gamma$ decreases from $\sim 1$ to
$\sim 0.5$, after the core has become supercritical
(which makes bar growth possible) but before a rapid collapse
sets in (which leaves little time for the bar to grow). The
bar growth removes an oft-invoked objection against the standard
scenario of ambipolar diffusion-driven core formation based
on the elongated or filamentary shape of dense cores
\citep{DWard-Thompson99}. Indeed, the triaxial nature of the
dense cores, coupled with the relatively slow, subsonic infall
motion inferred  \citep{CLee01}, may provide the strongest 
support yet for the standard scenario.

\begin{figure*}
\plotone{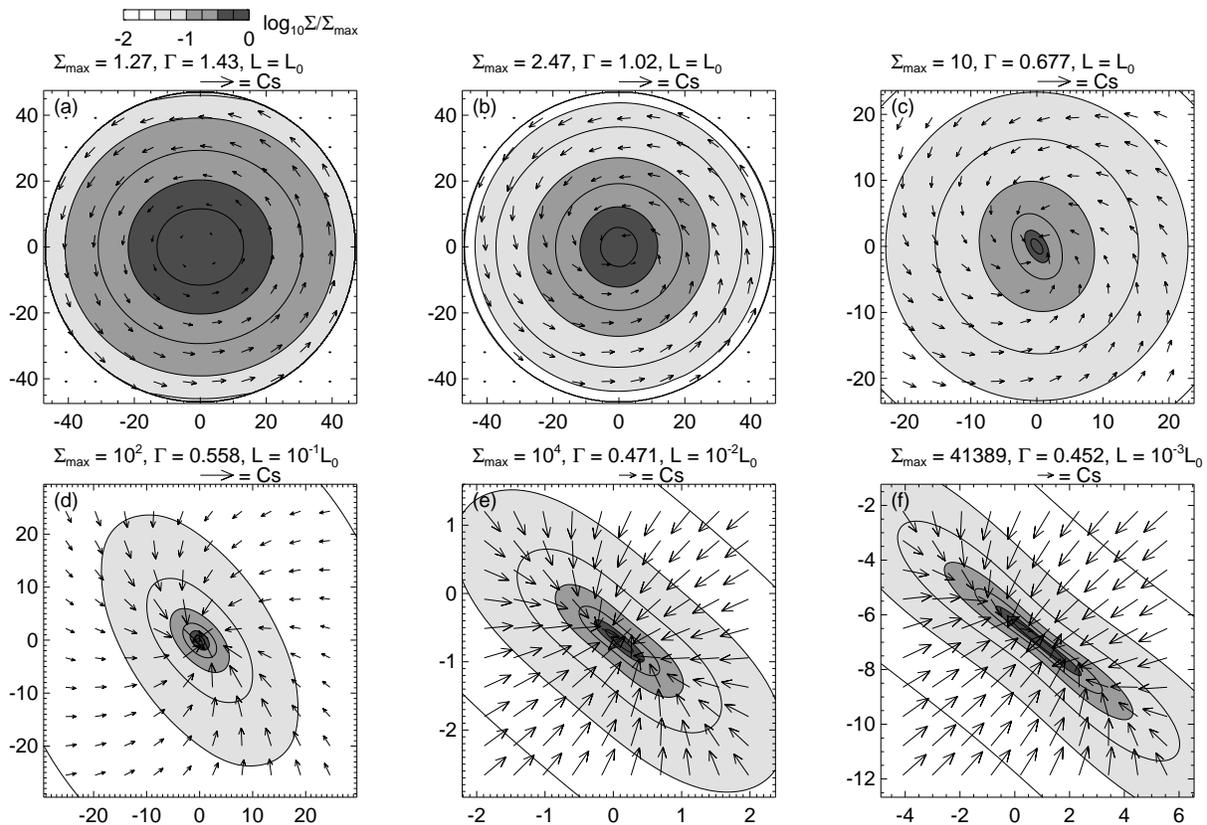}
\caption{
Surface density and velocity distributions for the model with
$(\Gamma_0, r_0, \omega)= (1.5, \, 7.5\pi, \, 0.1)$
at six dimensionless times: (a) t=0, (b) 17.49, (c) 24.46, (d) 25.21, (e)
25.2500, and (f) 25.2503.
In panels (c) through (f), only the central regions are shown.
The contours in each panel are for surface density normalized by
the maximum, with values indicated by the grayscale.
The numbers above each panel are the maximum surface density
($\Sigma_{\rm max}$), flux-to-mass ratio ($\Gamma$) at the
density peak, and length unit for each panel. The arrows
show the velocity vectors and are normalized by the arrow above
each panel. If we choose $\Sigma_{0,\rm ref} = 0.01$ g cm$^{-3}$,
$T_{\rm eff} = 30$ K, then the units for time, length and speed would
be $t_0 = 2.46\times 10^5$ yr, $L_0=0.082$ pc and $c_s=0.33$~km~s$^{-1}$.
The dimensionless flux-to-mass ratios averaged in the darkest regions
of the panels are (a) 4.46, (b) 2.77, (c) 1.26, (d) 0.645, (e) 0.503, 
and (f) 0.480, respectively. Note that the ``bar'' as defined by the
darkest region becomes completely supercritical at late times, even 
though the cloud as a whole remains subcritical. The maximum 
volume densities in panels (a) through (f) would be (a) $4.46\times 10^3$
cm$^{-3}$, (b) $1.70\times 10^4$ cm$^{-3}$, (c)
$2.79\times 10^5$ cm$^{-3}$ (d) $2.80\times 10^7$  cm$^{-3}$,
(e) $2.78\times 10^{11}$  cm$^{-3}$, and (f) $4.77\times 10^{12}$
cm$^{-3}$, respectively.
\label{fig:1}}
\end{figure*}

The continued evolution of bar-like cores, such as the one shown
in Figure \ref{fig:1}c, would formally lead to a singular filament,
if the isothermal assumption is kept. However, it is well known
that the equation of state stiffens when the optical depth exceeds
unity \citep[see however][]{HMasunaga99}. The stiffening
slows down the collapse, allowing more time for the elongated bar
to fragment gravitationally into small pieces. In fact, 
the major axis of the bar exceeds twice the critical wavelength
for fragmentation of an infinitely-long filament 
\citep{RLarson85} in panels (e) and (f) of Figure \ref{fig:1}.
Such bar formation is also shown by \citet{TMatsumoto99} and 
\citet{LSigalotti01} for nonmagnetized clouds.
Recent examples of (nonmagnetic) bar fragmentation
induced by the stiffening of
equation of state are given in \citet{ABoss00b}
and \citet{LSigalotti01}. 
For magnetized
bars, one needs to consider in addition the rapid decoupling of
magnetic field from matter, which occurs above a density of order
$10^{10}$~cm$^{-3}$ \citep{RNishi91}. The
decoupling decreases the magnetic support quickly, and is in
some sense equivalent to a sudden cooling. It should make the
fragmentation easier. 

\citet{ABoss00} studied the fragmentation of 3D magnetic clouds numerically,
treating the magnetic forces and ambipolar diffusion in an approximate
way. He concluded that magnetic fields can enhance cloud fragmentation,
by reducing the tendency for the development of a central singularity,
which would make fragmentation more difficult. 
We also find
that magnetic fields can promote fragmentation, but for a different
reason. Strong magnetic fields can support clouds of more
than one Jeans mass, which provides an initial condition that is
conducive to fragmentation once the magnetic support weakens. We
believe that it is the multi-Jeans mass nature of the magnetically
subcritical clouds that drives the nonlinear growth of the bar (and
higher order) mode. We conclude that the magnetically subcritical clouds
envisoned in the standard scenario of star formation can produce not 
only single stars but also, perhaps even preferentially, binary and 
multiple star systems. A parameter survey will be carried out to firm 
up this conclusion.



\end{document}